\documentclass[namedreferences]{SolarPhysics}
\usepackage[optionalrh]{spr-sola-addons} % For Solar Physics 
\usepackage{epsfig}          % For eps figures, old commands
\usepackage{graphicx}        % For eps figures, newer & more powerfull
\usepackage{color}           % For color text: \color command
\usepackage{url}             % For breaking URLs easily trough lines
\newcommand{\etal}{{\it et al.}}

\newcommand{\adv}{    {\it Adv. Spa. Res.}}

\newcommand{\aap}{    {\it Astron. Astrophys.}}

\newcommand{\aapr}{   {\it Astron. Astrophys. Rev.}}

\newcommand{\apj}{    {\it Astrophys. J.}}

\newcommand{\mnras}{  {\it Mon. Not. Roy. Astron. Soc.}}
\newcommand{\nat}{    {\it Nature}}
\newcommand{\pasp}{   {\it Pub. Astron. Soc. Pac.}}

\newcommand{\solphys}{{\it Solar Phys.}}

\newcommand{\ssr}{    {\it Space Sci. Rev.}}

\begin{document}

\begin{article}

\begin{opening}

\title{Perspectives in Global Helioseismology, and the Road
  Ahead\footnote{Invited Review}}

\author{William~J.~\surname{Chaplin}$^{1}$\sep
        Sarbani~\surname{Basu}$^{2}$\sep
       }
\runningauthor{W. J. Chaplin, S. Basu}
\runningtitle{Perspectives in global helioseismology}

   \institute{$^{1}$ School of Physics and Astronomy, University of
                     Birmingham, Edgbaston, Birmingham B15 2TT, UK
                     email: \url{w.j.chaplin@bham.ac.uk}\\ 
              $^{2}$ Department of Astronomy, Yale University,
                     P.O. Box 208101, New Haven, CT 06520-8101, USA
                     email: \url{sarbani.basu@yale.edu} \\
             }

\begin{abstract}

We review the impact of global helioseismology on key questions
concerning the internal structure and dynamics of the Sun, and
consider the exciting challenges the field faces as it enters a fourth
decade of science exploitation. We do so with an eye on the past,
looking at the perspectives global helioseismology offered in its
earlier phases, in particular the mid-to-late 1970s and the 1980s.  We
look at how modern, higher-quality, longer datasets coupled with new
developments in analysis, have altered, refined, and changed some of
those perspectives, and opened others that were not previously
available for study. We finish by discussing outstanding challenges
and questions for the field.

\end{abstract}


\end{opening}

  \section{Introduction}
  \label{sec:intro}

The field of global helioseismology -- the use of accurate and precise
observations of the globally coherent modes of oscillation of the Sun
to make inference on the internal structure and dynamics of our star
-- is about to enter its fourth decade. The observational starting
point for global helioseismology was marked in the mid-to-late 1970s
by several key papers: First, the observational confirmation by
Deubner (1975), and independently by Rhodes, Ulrich and Simon (1977),
of the standing-wave nature of the five-minute oscillations observed
on the surface of the Sun, which was proposed by Ulrich (1970), and
Leibacher and Stein (1971); and then the discovery that the
oscillations displayed by the Sun were truly global whole-Sun,
core-penetrating, radial-mode pulsations (Claverie \etal,
1979). Previous observations of pulsating stars had revealed many
objects that were oscillating in one, or at most a few, modes. The
rich spectrum of oscillations displayed by the Sun was a different
matter entirely. Christensen-Dalsgaard and Gough (1976) pointed out
the great potential that a multi-modal spectrum could offer: the
information content of the observations could potentially be so great
as to allow a reconstruction of the internal structure of the
star. More than 30 years on, exquisite observational reconstructions
of the internal structure and dynamics of the Sun are in everyday use,
thanks to helioseismology.

Several thousand modes of oscillation of the Sun have to date been
observed, identified, and studied. The oscillations are standing
acoustic waves, for which the gradient of pressure ($p$) is the
principal restoring force. The modes are excited stochastically, and
damped intrinsically, by the turbulence in the outermost layers of the
sub-surface convection zone. The stochastic excitation mechanism
limits the amplitudes of the $p$ modes to intrinsically weak
values. However, it gives rise to an extremely rich spectrum of modes,
the most prominent generally being high-order overtones. Detection of
individual gravity modes -- for which buoyancy acts as the principal
restoring force -- remains an important goal for the field. The $g$
modes are confined in cavities in the radiative interior, and if
observed would provide a much more sensitive probe of conditions in
the core than the $p$ modes.

The small-amplitude solar oscillations may be described in terms of
spherical harmonics $Y^{m}_{\ell}(\theta,\phi)$, where $\theta$ and
$\phi$ are the co-latitude and longitude respectively:
  \begin{equation}
  Y^{m}_{\ell}(\theta,\phi) = (-1)^m c_{\ell m} P^{m}_{\ell}(\cos
  \theta) \exp({\rm i} m\phi).
  \label{eq:ylm}
  \end{equation}
In the above, the $P^m_\ell$ is a Legendre function, and the $c_{\ell
m}$ is a normalization constant. The $p$ modes probe different
interior volumes, with the radial and other low-degree (low-$\ell$)
modes probing as deeply as the core. This differential penetration of
the modes allows the internal structure and dynamics to be inferred,
as a function of position, to high levels of precision not usually
encountered in astrophysics. The Sun has not surprisingly been the
exemplar for the development of seismic methods for probing stellar
interiors.  Extension of the observations to other Sun-like stars
(asteroseismology) has demonstrated that the Sun-like oscillations are
a ubiquitous feature of stars with sub-surface convection zones.

In this review our aim is to look at some of the recent advances that
global helioseismology has made for studies of various aspects of the
internal structure of the Sun. We discuss the observational challenge
posed by the detection and identification of individual $g$ modes. We
also seek to provide in each section some historical context for
discussion of the contemporary results and challenges. We round out
the review by looking to the future, in particular the need for
continued multi-instrument, multi-network observations of the global
modes; and we finish by listing some important questions and
challenges for global helioseismology.

  \section{The Standard Solar Model and the Abundance Problem}
  \label{sec:models}

The first important inference to be made by helioseismology on the
internal structure concerned the depth of the solar convection
zone. Gough (1977) and (independently) Ulrich and Rhodes (1977)
realized that a mismatch between computed $p$-mode frequencies (Ando and
Osaki, 1975) and the observed frequencies (actually the locations in
frequency of ridges in the $k$-$\omega$ diagram) could be reconciled
by increasing the depth of the convection zone by about 50\% compared
to typical values used at the time.

Investigations soon followed into the compatibility of solar models,
having different heavy-element abundances, with the observed $p$-mode
frequencies. An important aim was to see whether models having low
initial heavy element abundances, but significant accretion rates,
could reconcile the ``solar neutrino problem'' and at the same time be
consistent with the seismic data (Christensen-Dalsgaard, Gough, and
Morgan, 1979). The only seismic frequencies that were available when
this work was done were those for modes that penetrated the
near-surface layers. As noted above, these seismic data favoured a
deep convection zone. This meant that the seismic data were also at
odds with a low surface abundance of helium: a deeper zone goes hand
in hand with a higher helium abundance. Given that the amounts of
helium and heavier elements are inexorably tied together, this seemed
to suggest the heavy element abundance could not be low as well.

More robust conclusions were possible once frequencies on the
core-penetrating, low-$\ell$ modes became available, from the early
1980s onwards (\textit{e.g.}, see Christensen-Dalsgaard and Gough,
1980; 1981). Once the Kitt Peak data (Duvall and Harvey, 1983) had
``filled the gap'' between the high-$\ell$ and low-$\ell$ data that
were already available, inversions for the internal sound speed were
possible.  These data, and more modern data, have resulted in detailed
inference on the solar structure. For instance, the position of the
convection zone (\textit{e.g.}, Basu and Antia, 1997) and the
convection-zone helium abundance (\textit{e.g.}, D\"appen \etal, 1991;
Basu and Antia, 1995) are both known to high precision. These
inferences on the solar structure, and the ability to determine
sound-speed and density differences between solar models and the Sun,
provide a means to test how well solar models fare against the Sun,
and thereby allows tests of the input physics to be made. For example,
the inversions of Christensen-Dalsgaard \etal\, (1985) suggested there
were problems with computation of the astrophysical
opacities. Significant improvements were made in the OPAL opacity
tables (\textit{e.g.}, Iglesias and Rogers, 1991; 1996). Further
improvements to the standard models followed with the routine
inclusion of diffusion and settling (\textit{e.g.}, Demarque and
Guenther, 1998; Cox, Guzik, and Kidman, 1989; Christensen-Dalsgaard,
Proffitt, and Thompson, 1993). Tests are also possible for the
equation of state (Christensen-Dalsgaard and D\"appen, 1992; Basu and
Christensen-Dalsgaard, 1997; Elliott and Kosovichev, 1998), and as a
result of the improvements that followed most modern solar models are
now constructed with the OPAL2001 equation of state (Rogers and
Nayfonov, 2002).

Early inversions, such as those by Christensen-Dalsgaard \etal\,
(1985), appeared to rule out the possibility of a low helium and low
heavy-element abundance. While the issue of the solar helium abundance
has been settled because it can be inferred from the very precise
$p$-mode frequencies, the heavy-element abundance question is live
again, and it may be said that the field is confronted with a ``solar
abundance problem''.  Whereas in the late 1970s and early 1980s, the
question of which abundances fitted the helioseismic data best was an
open question -- the helioseismic data were new, and the community was
still feeling its way on fully exploiting the data -- today we know
the answer to that question. The issue is instead very much open
because estimates of the photospheric abundances, provided by
spectroscopy, have been revised downwards.

\begin{figure}

 \centerline{\hbox{\includegraphics[width = 170pt]{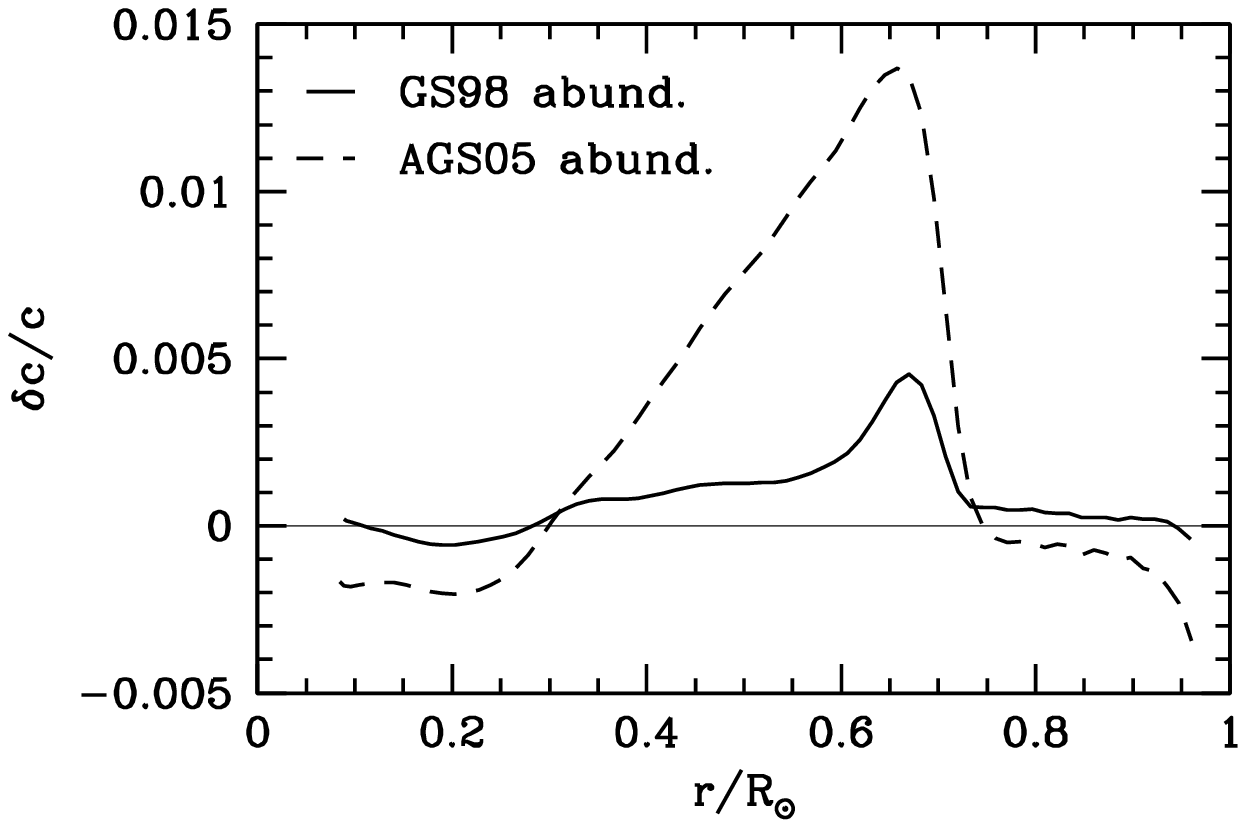}\quad 
                   \includegraphics[width = 170pt]{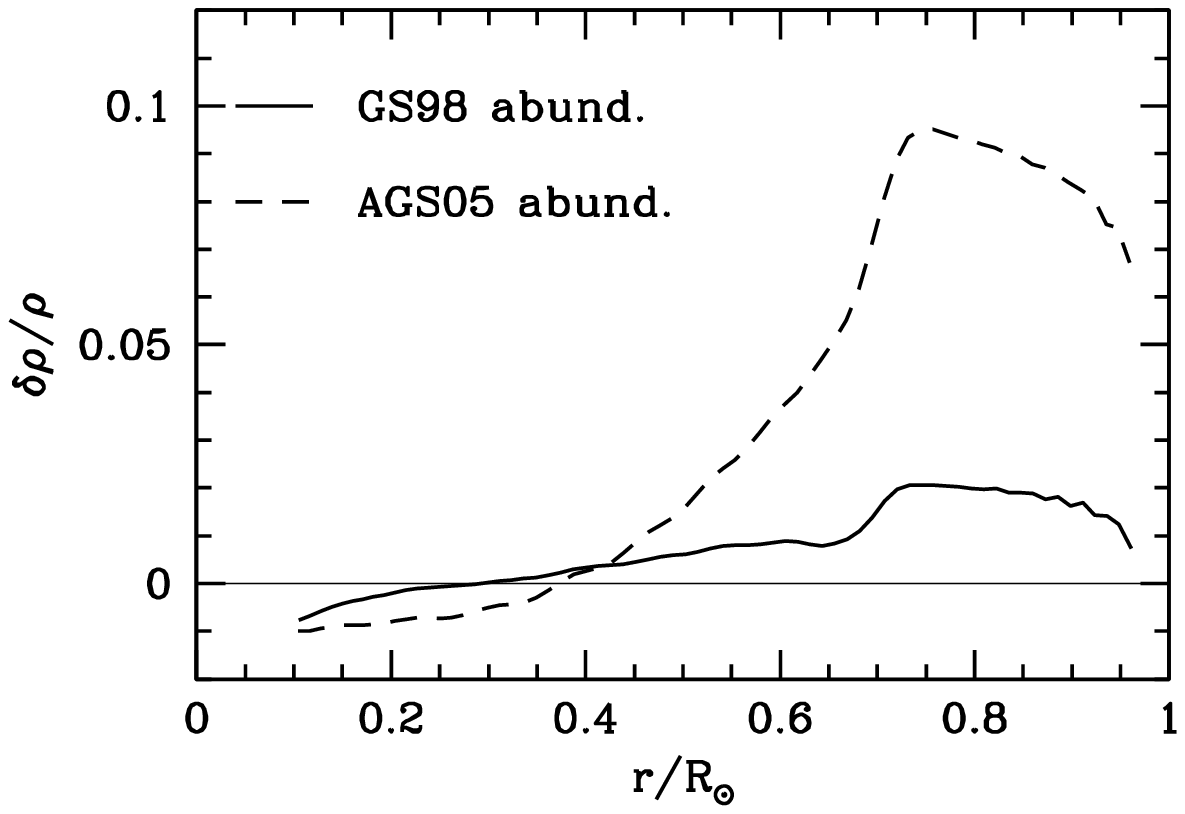}}}

 \caption{The relative sound-speed difference (panel a) and density
  difference (panel b) between the Sun and a standard solar model
  constructed with the GS98 metallicity [model BP04(Garching)], and
  also a standard solar model constructed with the AGS05 metallicity
  [model BS05(AGS,OPAL)] of Bahcall, Basu, and Serenelli, 2005. The
  model with GS98 $Z/X$ has a CZ He abundance of $Y_{\rm CZ}=0.243$
  and CZ base at $R_{\rm CZ}=0.715\,r/R_\odot$. The AGS05 model has
  $Y_{\rm CZ}=0.230$ and $R_{\rm CZ}=0.729\,r/R_\odot$.}

  \label{fig:csq}

\end{figure}

One of the important inputs into these models is the heavy-element
abundance ($Z$) or alternatively, the ratio of the heavy element to
the hydrogen abundance ($Z/X$). Solar models that have shown a good
agreement with results from helioseismology were constructed with the
solar abundance as given by Grevesse and Noels (1993), or more
recently by Grevesse and Sauval (1998; henceforth GS98). The GS98
table shows that $Z/X=0.0229$, \textit{i.e.}, $Z=0.0181$ for the
Sun. The situation has changed recently. Asplund \etal\, (2000, 2004),
Allende Prieto, Lambert, and Asplund (2001, 2002) and Asplund \etal\,
(2005), find that the solar heavy-element abundances need to be
reduced drastically, based on what they claim are better calculations
with improved models of the solar photosphere. This lead Asplund,
Grevesse, and Sauval (2005; henceforth AGS05) to compile a table of
solar abundances, with $Z/X=0.0166$ (\textit{i.e.}, $Z=0.0122$). This
has resulted in considerable discussion in the community since the
sound-speed and density profiles of models constructed with AGS05 do
not agree well with the Sun. This disagreement can be seen in
Figure~\ref{fig:csq} where we show the density and sound-speed
differences between the Sun and two solar models, one constructed with
the GS98 abundances and the other with the AGS05 abundances.

%%%%%%%%%%%%%%%%%%%%%%%%%%%%%%%%%%%%%%%%%%%%%%%%%%%%%%%%%%%%%%%%%%%%%%%%%%%

%%%%%%%%%%%%%%%%%%%%%%%%%%%%%%%%%%%%%%%%%%%%%%%%%%%%%%%%%%%%%%%%%%%%%%%%%%%

The mismatch between the models with low $Z$ and the Sun is most
striking in the outer parts of the radiative interior, a result of the
fact that the low-$Z$ models have a much shallower convection zone
than the Sun.  There are however, differences in other regions too:
all standard models with AGS05 abundances have low helium abundance in
the convection zone (\textit{e.g.}, Montalb\'an \etal,~2004; Guzik,
Watson, and Cox, 2005; Bahcall, Basu, and Serenelli, 2005); the
seismic signatures of the ionization zones do not match observations
(Lin, Antia, and Basu, 2007); and the helioseismic signatures from the
core do not match observations either (Basu \etal, 2007).

There have been several attempts to reconcile low-$Z$ solar models
with helioseismic data. Given that the largest discrepancy is at the
base of the convection zone, the first attempts involved modifying the
input opacities. It was found that large changes in opacity, in the
range 11\% to 21\%, would be needed at temperatures relevant to the
base of the convection zone to resolve the problem (Montalb\'an \etal,
2004; Basu and Antia, 2004; Bahcall \etal,~2005).  However, later
re-calculation of the opacities by the OP group (Badnell \etal, 2005)
showed an opacity increase of only 2\%. Other attempts included
increasing the diffusion coefficient (\textit{e.g.}, Montalb\'an
\etal,~2004; Basu and Antia, 2004; Guzik \etal,~2005). A large change
was needed to get the correct position of the convection-zone base,
which resulted in an extremely low convection-zone helium
abundance. Attempts were also made to increase the metallicity of the
models by increasing the abundance of uncertain elements such as Neon,
which does not have any photospheric lines (Antia, and Basu, 2005;
Bahcall \etal,~2005b). However it is not clear if such an increase is
justified in the case of the Sun (Schmelz \etal,~2005; Young
2005). Other attempts involve \textit{ad hoc} prescriptions of mixing
at the tachocline (Turck-Chi\`eze \etal, 2004; Montalb\'an \etal,
2006) or mixing by gravity waves (Young and Arnett, 2005).  Late
accretion of low-$Z$ material by the zone has been tried as well
(Guzik \etal,~2005; Castro, Vauclair, and Richard, 2007).  None of the
models match the Sun unless two or more modifications are used
(Montalb\'an \etal,~2004) and even in those cases the changes in
physics have to be fine-tuned carefully.

Given that attempts to adjust physical inputs in an \textit{ad hoc}
manner have not resulted in low-$Z$ solar models that agree with the
Sun, Antia and Basu (2006) tried to derive $Z$ for the Sun using
signatures of the heavy-element ionization zones. The method was
similar to that used by Basu and Antia (1995) to determine the He
abundance of the Sun. They found a solar $Z$ of $0.0172\pm 0.002$, a
value close to the GS98 value and much larger than the AGS05
value. The errors in the result are not affected by errors in
opacity. While the Antia and Basu (2006) result was obtained from
helioseismic signatures in the upper convection zone, using data from
the solar core Chaplin \etal\, (2007a) also concluded that $Z$ in the
Sun has to be high.  They found that the mean molecular weight
averaged over the inner 20\% by radius of the Sun is in the range
$0.7209$ to $0.7231$ and that the corresponding surface $Z$ is in the
range $0.0187$ to $0.0239$.

The obvious discrepancy between the low-$Z$ models and the Sun creates
a problem that has not yet been resolved. If the new, lower abundances
are correct then the obvious culprit is missing or incorrect physics
in the solar models. If however, the lower abundances are incorrect,
then we can be confident that the input physics in our models is
within errors. This supposition is supported by the seismically
determined value of $Z$. The seismic $Z$ determinations have been
achieved using techniques that depend on different inputs, and despite
the differences in techniques and dependencies on different inputs, all
seismic estimates of $Z/X$ are consistent with the higher GS98
abundances, and they agree with each other as well.

If the convection-zone abundances of the Sun are indeed consistent
with the low abundances compiled by AGS05, then almost all the input
physics that goes into construction of stellar models must be much
more uncertain than has been assumed to be the case. It is also
possible that some fundamental process is missing in the theory of
stellar structure and evolution, but it is difficult to speculate what
that could be. On the other hand, if the GS98 abundances are correct
then the currently known input physics is consistent with seismic
data, and the AGS05 abundances need to be revised upwards. It is
easier to list reasons why the new abundances could be
incorrect. These include the fact that the 3D convection simulation
used in the atmospheric models may not have the correct thermal
structure (Ayres, Plymate, and Keller, 2006). There may be some
effects from having a grid of finite resolution: Scott \etal\, (2006)
found significant changes in the line bisectors high in the atmosphere
as they changed their resolution. In addition there could be problems
with the non-LTE effects used in the line-formation calculations and
atomic physics.

In conclusion, disagreements between helioseismic estimates and recent
spectroscopic estimates of the solar heavy-element abundance call for
continuation of the careful examination of solar atmospheric models,
and also models of the solar interior.

  \section{The Internal Rotation and Dynamics}
  \label{sec:rot}

The time-averaged internal rotation profile revealed by global
helioseismology has in many respects been something of a surprise (see
Thompson \etal, 2003 for an excellent review on the internal
rotation). The known pattern of differential rotation at the surface
was observed to penetrate the interior, down to the base of the
convection zone, but not in the manner expected. The seismic data then
revealed the solar tachocline, a narrow region in the stably
stratified layer just beneath the base of the convective envelope,
which mediates the transition from differential rotation above to a
solid-body-like profile in the radiative zone below. This
solid-body-like profile, with its ``slow'' (\textit{i.e.}, surface-like)
rotation rate, was of course the other surprise.

The paradigm for the spin-down of Sun-like stars involves the action
of a dynamo in the outer envelope. Magnetic braking slows the rate of
rotation in the envelope. The question then arises as to the degree of
coupling between the radiative interior and the envelope. Coupling
allows the envelope to draw on the large reservoir of angular momentum
residing in the core. This has two consequences. First, it will delay
the rate at which the envelope is spun down. And second, it will bleed
momentum from the core, thereby altering the rotation in the deep
interior. The extent to which the core and envelope are coupled
therefore plays an important r\^ole in the dynamic evolution of a
star. In the pre-helioseismic era, the conventional wisdom was that
the core and deep radiative interior of the Sun would be expected to
rotate much more rapidly than the layers above.

Further to answering questions posed by stellar evolution theory,
models and conjectures on the rotation were also of considerable
interest for attempts to constrain one of the well-known tests of
Einstein's general theory of relativity, the advance in the perihelion
of Mercury test.  Observations of the surface oblateness of the Sun,
by Dicke and Goldenberg (1967), had suggested that the solar
gravitational quadrupole moment was large enough to give a significant
contribution to the perihelion advance. This created something of a
problem, for it reduced the fraction of the contribution that could be
set aside as being relativistic in origin to the point where what was
left over conflicted with the prediction for the general theory of
relativity.  Interest in competing theories of relativity was
reinvigorated (\textit{e.g.}, the theory of Brans and Dicke (1961),
which could be ``tuned'' into agreement with the oblateness
observations).

The result of Dicke and Goldenberg had important implications for the
Sun's interior structure, for it suggested a rapidly spinning core
might be needed to account for the apparently large surface
oblateness. The concept of rapid internal rotation also happened to be
in vogue at the time for another reason: it offered a possible way to
solve the solar neutrino problem.  In the presence of rapid internal
rotation thermal pressure would no longer be required to carry the
full burden of support to maintain the star in equilibrium, meaning
temperatures in the core could be lower than previously thought. A
nice snapshot is provided by three papers: Ulrich (1969); Demarque,
Mengel, and Sweigart (1973); and Roxburgh (1974).

What of the outer layers?  Pre-helioseismic models of rotation in the
convection zone gave a pattern in which the rotation was constant on
cylinders wrapped around the rotation axis (\textit{e.g.}, Glatzmaier,
1985; Gilman and Miller, 1986). Small-diameter cylinders intersect the
solar surface at high latitudes, while larger cylinders do so at low
latitudes, in the vicinity of the solar equator. In order to match to
the surface differential rotation, material lying on the surface of a
small cylinder must rotate less rapidly than plasma on a larger
cylinder.  An important consequence of the rotation models was that
they therefore predicted an increase of rotation rate with increasing
radius.

 \subsection{Observed Rotation: the Deep Interior}
 \label{sec:deep}

Helioseismic inference on the rotation of the deep interior demanded
estimates of the rotational frequency splittings of the
core-penetrating low-$\ell$ $p$ modes. The first estimates of the
rotational frequency splittings (Claverie \etal, 1981) suggested the
core might indeed be spinning more rapidly than the outer layers, but
by nowhere near enough to give the surface oblateness claimed by Dicke
and Goldenberg.  The first inversion for the internal rotation (Duvall
\etal, 1984) showed that the outer parts of the radiative interior were
actually rotating at a rate not dissimilar to the surface, a result
that all but ruled out the possibility of a significant gravitational
quadrupole moment.

The intervening years have seen a steady, downward revision of the
magnitudes of the quoted estimates of the low-$\ell$ rotational
frequency splittings (\textit{e.g.}, see discussion in Chaplin,
2004). By the mid 1990s this downward trend had flattened out. Alas,
the trend was not solar in origin. It was the result of having longer,
higher-quality datasets available, coupled to a better understanding
of the subtleties and pitfalls involved in extracting the frequency
splittings (\textit{e.g.}, Appourchaux \etal, 2000a; Chaplin \etal,
2001a). In short: estimates from short datasets tend to overestimate
the true splittings because there is insufficient resolution in
frequency to properly resolve individual components (which is why the
initial estimates of Claverie \etal, were high).  This is surely one of
the best examples helioseismology can offer on how accumulation of
data from long-term observations (coupled with a better feel for the
analysis) can give significant improvement on accuracy of inference on
the internal structure.

Inversions made with the modern, high-quality data (\textit{e.g.},
Eff-Darwich, Korzennik, and Jim\'enez-Reyes, 2002; Couvidat \etal,
2003; Garc\'ia \etal, 2004) give well-constrained estimates on the
rotation down to $r/R_{\odot} \simeq 0.25$, where the rotation rate is
observed to be similar to that in the mid-latitude near-surface
layers. We comment below in Section~\ref{sec:howrot} on how the
solid-body-like rotation might be enforced in the radiative interior.

The conclusion that the quadrupole moment is of insufficient size to
give a significant contribution to the perihelion advance of Mercury
has been upheld by the modern observations of slow rotation in the
interior (\textit{e.g.}, Pijpers, 1998; Roxburgh, 2001). Contemporary
measurements of the solar shape -- made, for example, with MDI data
(Emilio \etal, 2007) -- show a minute oblateness. The possibility of
rapid internal rotation providing a solution to the solar neutrino
problem was therefore moot, not only because of the slow rotation, but
also because agreement between the sound-speed profiles of solar
models and the Sun showed the problem was not one in solar physics
(\textit{e.g.}, Bahcall \etal, 1997), a result confirmed by
observations made by the Sudbury Neutrino Observatory (\textit{e.g.},
Ahmad \etal, 2001; 2002; Ahmed \etal, 2004).

But what of the rotation rate in the core itself? This remains very
uncertain. Use of the $p$ modes presents several difficulties: only a
small number of the modes penetrate the core, and those that do have
only a modest sensitivity to the rotation.  It will be through the
measurement of the rotational frequency splittings of $g$ modes that a
clear picture of the rotation in the core will properly emerge. Mathur
\etal\, (2007) have demonstrated that by augmenting the $p$-mode
splittings with splittings of a small number of $g$ modes, it should
be possible to obtain precise, and reasonably accurate, estimates of
the rotation profile throughout a substantial fraction of the
core. This is surely sufficient reason alone to redouble our effects
to detect individual $g$ modes. We discuss the current status of the
observational claims in Section~\ref{sec:gmodes} below.

 \subsection{Observed Rotation: the Convection Zone and Tachocline}
 \label{sec:contach}

Initial glimpses of the rotation in the near-surface layers were
provided by Rhodes, Ulrich, and Deubner (1979). However, it was Brown
(1985) who presented the first evidence that demonstrated the surface
differential rotation penetrated the convection zone. Further studies,
using inversion of the rotational frequency splittings, followed
(\textit{e.g.}, Brown, and Morrow, 1987; Kosovichev, 1988; Brown
\etal, 1989; Dziembowksi, Goode, and Libbrecht, 1989; Rhodes \etal,
1990; Thompson, 1990). By the end of the 1980s, the rotation
inversions were able to show that the differential rotation underwent
a marked transition at the base of the convection zone to something
resembling a solid-body-like profile below (Christensen-Dalsgaard and
Schou, 1988). The tachocline (Speigel and Zahn, 1992) had been
discovered.

Analysis with the more extensive modern data (\textit{e.g.}, Antia and
Basu, 2003) indicates that the characteristic thickness of the
tachocline is only a few per cent of the solar radius. The tachocline
is oblate, and slightly thicker at the solar equator. The steep
gradient in rotation present across the tachocline -- much stronger
than anything present elsewhere in the outer layers -- means it is of
considerable interest to the dynamo modelers, and is an attractive
site in which to locate stretching, and winding-up, of magnetic field
(poloidial to toroidal) by the $\Omega$ effect (\textit{e.g.}, see
Tobias, 2002).

\begin{figure*}
 \centerline
% {\includegraphics[width=0.8\textwidth,clip=]{hemi.ps}}
 {\includegraphics[width=0.8\textwidth,clip=]{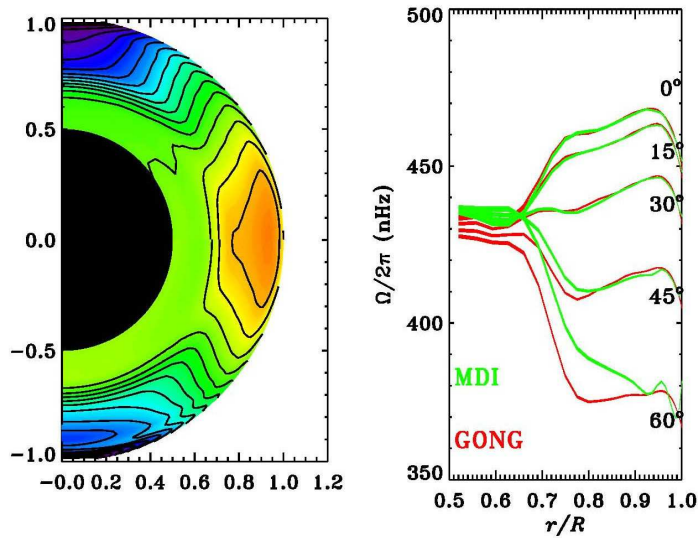}}

 \caption{The mean, time-averaged rotation profile in the convection
 zone and outer parts of the radiative zone. Left-hand panel: 2D
 cutaway, showing the mean profile obtained from GONG (upper half) and
 MDI (lower half) observations. Right-hand panel: mean rotation
 profile at different latitudes. (Courtesy of R.~Howe.)}

 \label{fig:meanrot}
\end{figure*}

What of the rotation in the convection zone itself? The pattern
revealed by analysis of the modern data (Figure~\ref{fig:meanrot})
does not match the rotation-on-cylinders prediction. Rather, the
rotation is approximately constant on lines inclined some $27^{\rm o}$
to the rotation axis (\textit{e.g.}, Gilman and Howe,
2003). Furthermore, the rotation rate decreases with radius in the
low- to mid-latitude layers very close to the surface (\textit{e.g.},
Corbard and Thompson, 2002). While there is a general consensus that
the differential rotation in the convection zone is driven by thermal
perturbations the challenge remains to understand in detail the mean
observed profile (\textit{e.g.}, see Rempel, 2005; Miesch, Brun, and
Toomre, 2006).

%%%%%%%%%%%%%%%%%%%%%%%%%%%%%%%%%%%%%%%%%%%%%%%%%%%%%%%%%%%%%%%%%%%%%%%%%%%%%%%

%%%%%%%%%%%%%%%%%%%%%%%%%%%%%%%%%%%%%%%%%%%%%%%%%%%%%%%%%%%%%%%%%%%%%%%%%%%%%%%

Some of the most striking results of helioseismology have come from
the detection of small, but significant, temporal variations of the
rotation rate in the convection zone, which carry signatures of the
solar activity cycle. We shall discuss these variations in
Section~\ref{sec:cyc} below.

 \subsection{How is the Tachocline Confined, and Solid-Body Rotation Enforced?}
 \label{sec:howrot}

The existence of the tachocline has raised several fundamental
questions regarding the dynamic evolution of the Sun. This thin layer
matches the transition in rotational behaviour above and below, and
must mediate or act as the intermediary for the transfer of angular
momentum from the immense reservoir in the core to the outer envelope
and beyond, as the star evolves.  In order for the rotation to change
its character, something must be acting in the radiative interior to
mix angular momentum in latitude, so that the differential rotation
from the convective zone above is smoothed out, or removed, below. The
mechanism must be anisotropic, in the sense that it must be much less
efficient at mixing angular momentum in the radial as opposed to the
latitudinal direction in order to explain the narrow width of the
tachocline. Mixing by anisotropic turbulence is one
possibility. Another possibility is the effect of a fossil magnetic
field threading the radiative interior.  Gough and McIntyre (1998)
considered circulations that penetrate from the convection zone into
the tachocline, which are then diverted by a weak magnetic field
further down. The magnetic field acts to prevent the tachocline
spreading out in radius, in effect forming a firm lower boundary. The
pattern of rotation at low and high latitudes acts to keep field
``bottled up'' in the radiative interior. This magnetic field need
only have a strength that is a tiny fraction of that at the surface in
order to give the required effect. What is more, the magnetic field is
then also a prime candidate for enforcing the solid-body-like rotation
which is present in the radiative zone (see also Eggenberger, Maeder,
and Maynet, 2005). Brun and Zahn (2006) have also recently looked at
the problem of confinement of the tachocline by a magnetic
field. However, their results imply that a fossil field in the
radiative interior cannot prevent the radial spread of the tachocline,
and that, furthermore, it also cannot prevent penetration of the
differential rotation from the convection zone into the radiative
interior.

Another mechanism that has received attention as a possible means to
enforce solid-body rotation in the deep interior is angular momentum
transport by internal gravity (buoyancy) waves, which are excited at
the base of the convection zone. It has recently been demonstrated
(Charbonnel and Talon, 2005) that such models can in principle
redistribute angular momentum efficiently over time from the core to
the outer envelope. In the presence of shear turbulence, the gravity
waves also give rise to shear-layer oscillations that resemble the
``quasi-biennial'' oscillations observed in the Earth's atmosphere
(Talon, 2006).

  \section{The Changing Sun}
  \label{sec:cyc}

A rich, and diverse, body of observational data is now available on
temporal variations of the properties of the global $p$ modes.  The
signatures of these variations are correlated strongly with the
well-known 11-year cycle of surface activity, and as such the accepted
paradigm is that the ``seismic'' solar cycle is associated with
changes taking place in the outer layers, not the deep radiative
interior, of the Sun.

Evolutionary changes to the equilibrium structure of the Sun will of
course also leave their imprint on the $p$ modes, by virtue of a very
slow adjustment of the interior structure as the star ages sedately on
the main sequence. The frequencies of the low-$\ell$ $p$ modes are
predicted by the standard solar models to decrease by $\approx 1\,\rm
\mu Hz$ every $6 \times 10^6$\,years due to the evolutionary
effects. If we alter the timescale to something more practical from an
observer's point of view -- say ten years -- the evolutionary change
is reduced to only $\approx 10^{-6}\,\rm \mu Hz$. Measurement of such
tiny frequency changes, against the backdrop of variations due to the
solar cycle, and instrumental noise properties, is beyond the current
scope of the data. The observed variations of the $p$ modes, due to
changes in the outer layers, are some five orders of magnitude larger
than the predicted evolutionary variations.

The search for temporal variations of the properties of the $p$ modes
began in the early 1980s, following accumulation of several years of
global seismic data. The first positive result was uncovered by
Woodard and Noyes (1985), who found evidence in observations by ACRIM
for a systematic decrease of the frequencies of low-$\ell$ $p$ modes
between 1980 and 1984.  The first year coincided with high levels of
global surface activity, while during the latter period activity
levels were much lower. The modes appeared to be responding to the
Sun's 11-year cycle of magnetic activity.  The uncovered shifts were,
on average, about $0.4\,\rm \mu Hz$. This meant that the frequencies
of the most prominent modes had decreased by roughly 1 part in 10\,000
between the activity maximum and minimum of the cycle. By the late
1980s, an in-depth study of frequency variations of global $p$ modes,
observed in the Big Bear data, had demonstrated that the agent of
change was confined to the outer layers of the interior (Libbrecht and
Woodard, 1990).

The passage of time, and accumulation of data from the new networks
and instruments, has allowed us to study the frequency variations in
unprecedented detail, and to reveal signatures of subtle, structural
change in the sub-surface layers. It has led to the discovery of
solar-cycle variations in the mode parameters associated with the
excitation and damping (\textit{e.g.}, power, damping rate, and peak
asymmetry). Patterns of flow that penetrate a substantial fraction of
the convection zone have been uncovered as well as possibly (but
controversially) signatures of changes in the rotation rate of the
layers that straddle the tachocline. Let us say a little more about
these observations, and what they might mean for our understanding of
the solar variability.

 \subsection{Structural Changes}
 \label{sec:dstruct} 

%%%%%%%%%%%%%%%%%%%%%%%%%%%%%%%%%%%%%%%%%%%%%%%%%%%%%%%%%%%%%%%%%%%%%%%%%%%%%%%

\begin{figure*}
 \centerline
% {\includegraphics[width=0.7\textwidth,clip=]{fshifts_deck2k6.ps}}
 {\includegraphics[width=0.7\textwidth,clip=]{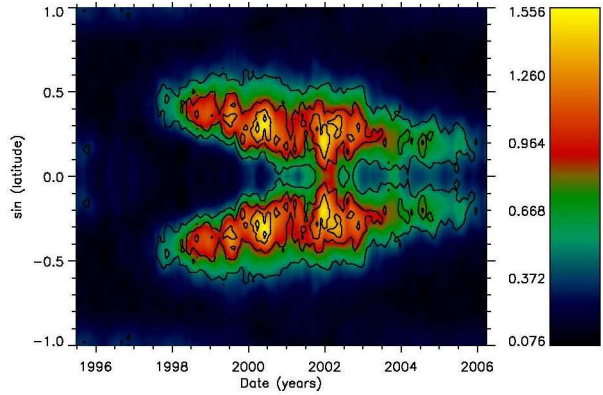}}

  \caption{Mode frequency shifts (in $\rm \mu Hz$) as a function of
   time and latitude. The values come from analysis of GONG data. The
   contour lines indicate the surface magnetic activity. (Figure
   courtesy of R.~Howe.)}

 \label{fig:numap}
\end{figure*}

%%%%%%%%%%%%%%%%%%%%%%%%%%%%%%%%%%%%%%%%%%%%%%%%%%%%%%%%%%%%%%%%%%%%%%%%%%%%%%%

%%%%%%%%%%%%%%%%%%%%%%%%%%%%%%%%%%%%%%%%%%%%%%%%%%%%%%%%%%%%%%%%%%%%%%%%%%%%%%%

\begin{figure*}
 \centerline
% {\includegraphics[width=0.7\textwidth,clip=]{wshifts_jun2k6.ps}}
 {\includegraphics[width=0.7\textwidth,clip=]{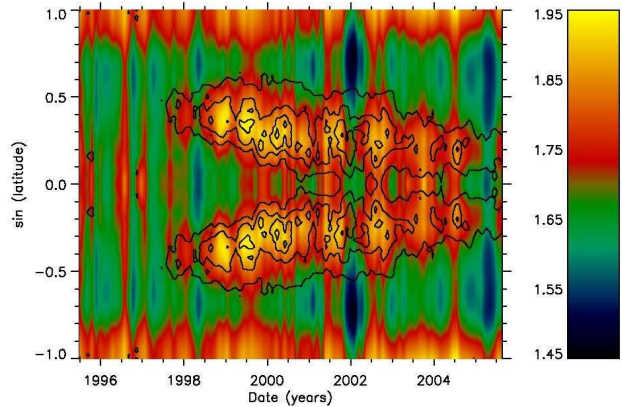}}

  \caption{Mode linewidth (in $\rm \mu Hz$) as a function of time and
   latitude. The values come from analysis of GONG data. The contour
   lines indicate the surface magnetic activity. (Courtesy of
   R.~Howe.)}

 \label{fig:widmap}
\end{figure*}

%%%%%%%%%%%%%%%%%%%%%%%%%%%%%%%%%%%%%%%%%%%%%%%%%%%%%%%%%%%%%%%%%%%%%%%%%%%%%%%

The modern seismic data give unprecedented precision on measurements
of the $p$-mode frequency shifts. From observations of the
medium-$\ell$ frequency shifts -- for example in GONG and MDI data --
it is possible to produce surface maps (Figure~\ref{fig:numap})
showing the strength of the solar-cycle shifts as a function of
latitude and time (Howe, Komm, and Hill, 2002). These maps bear a
striking resemblance to the butterfly diagrams that show variations in
the strength of the surface magnetic field over time. The implication
is that the frequency shift of a given mode depends on the strength of
that component of the surface magnetic field that has the same
spherical harmonic projection on the surface. This dependence is also
observed in studies of the frequency shifts of the less numerous
low-$\ell$ modes (see Chaplin, 2004, and references therein). The
precision in the medium-$\ell$ data is such that significant frequency
changes can now be tracked on timescales as short as nine days (see
Tripathy \etal, 2007). Meanwhile current results on frequency shifts
of high-$\ell$ modes ($\ell > 100$) -- extracted using global
helioseismology techniques (Rabello-Soares, Korzennik, and Schou,
2007) -- show trends that match to the medium-$\ell$ and low-$\ell$
shifts (\textit{e.g.}, Chaplin \etal\, 2001b).  In particular, when
the frequency shifts are multiplied by the mode inertia, and then
normalized by the inertia of a radial mode of the same frequency, the
modified shifts are found to be a function of frequency alone. The
high-$\ell$ modes provide important information since they are
confined in the layers close to the surface where the physical changes
responsible for the frequency shifts are also located.

Detailed comparison of the low-$\ell$ frequency shifts with changes in
various disc-averaged proxies of global surface activity provides
further tangible input to the solar cycle studies. This is because
different proxies show differing sensitivity to various components of
the surface activity.  Chaplin \etal\, (2007b) recently compared
frequency changes in 30\,years of BiSON data with variations in six
well-known activity proxies. Interestingly, they found that only
activity proxies having good sensitivity to the effects of
weak-component magnetic flux -- which is more widely distributed in
latitude than the strong flux in the active regions -- were able to
follow the frequency shifts consistently over the three cycles.

What is the physical mechanism behind the frequency shifts?  Broadly
speaking, the magnetic fields can affect the modes in two ways. They
can do so directly, by the action of the Lorentz force on the
plasma. This provides an additional restoring force, the result being
an increase of frequency, and the appearance of new modes. Magnetic
fields can also influence matters indirectly, by affecting the
physical properties in the mode cavities and, as a result, the
propagation of the acoustic waves within them. This indirect effect
can act both ways, to either increase or decrease the frequencies. The
exact nature of the physical changes is still somewhat controversial,
although Dziembowski and Goode (2005) have recently made important
headway on the problem. Their analysis of MDI $p$-mode frequency
shifts suggests it is indirect effects that dominate, in particular
changes to the near-surface stratification resulting from the
suppression of convection by the magnetic field.  They suggest that
the magnetic fields are too weak in the near-surface layers where the
$p$-mode shifts originate for the direct effect to contribute
significantly.

It is also interesting to note that Dziembowski and Goode found small,
but significant, departures for the lower-frequency $p$ modes of a
simple scaling of the frequency shifts with the inverse mode
inertia. The nature of these small departures suggests that there is a
contribution to the low-frequency shifts from deeper layers, due to
the direct effect of the magnetic fields. Similar departures in
behaviour had also been seen and noted by Chaplin \etal\, (2001b).

Variations of global $f$-mode frequencies reveal information on
changes in a thin layer which extends some 15\,Mm below the base of
the photosphere. The most recent observations (\textit{e.g.}, Lefebvre
and Kosovichev, 2005) suggest that as activity rises there is an
expansion between $r/R_{\odot} \sim 0.97$ and 0.99, and possibly a
contraction above $r/R_{\odot} \sim 0.99$. It is, however, not yet
possible to reconcile the observations with theoretical predictions of
the variations (see Lefebvre, Kosovichev, and Rozelot, 2007; also
Sofia \etal, 2005).

Variations very close to the surface, in the He~{\sc ii} ionization
zone at a depth $\approx 0.98\,r/R_{\odot}$, have also been revealed
by analysis of medium-$\ell$ $p$ modes. From appropriate combinations
of mode frequencies, Basu and Mandel (2004) uncovered apparent
solar-cycle variations in the amplitude of the depression in the
adiabatic index, $\Gamma_1$, in the He~{\sc ii} zone. These variations
presumably reflect the impact of the changing activity on the equation
of state of the gas in the layer. These results have since been
confirmed, using a different method to extract the acoustic signatures
of the He~{\sc ii} zone, and with only low-$\ell$ frequencies (Verner,
Chaplin, and Elsworth, 2006).

The results discussed above pertain to changes taking place very close
to the surface. What of possible changes deeper down? Chou and
Serebryanskiy (2005), and Serebryanskiy and Chou (2005), have found
intriguing evidence of signatures in the $p$-mode frequency shifts that
may reflect changes taking place near the base of the convection zone.
The authors suggest that the signatures they uncover are consistent
with a fractional perturbation to the sound speed, at depth 0.65 to
$0.67\,r/R_{\odot}$, of size a few parts in $10^5$ (assuming the
perturbation may be described as a Gaussian with a FWHM of
$0.05\,r/R_{\odot}$ in radius).

Surface maps, such as the frequency-shift map shown in
Figure~\ref{fig:numap}, may also be made for variations observed in
the mode powers and damping rates (Komm, Howe, and Hill, 2002), which,
like the frequency maps, show a close spatial and temporal
correspondence with the evolution of active-region field
(Figure~\ref{fig:widmap}). Meanwhile, peak asymmetry is the most
recent addition to the list of parameters that show solar-cycle
variations (Jim\'enez-Reyes \etal, 2007). Careful measurement of
variations in the powers, damping rates and peak asymmetries -- all
parameters associated with the excitation and damping -- allows
studies to be made of the impact of the solar cycle on the convection
properties in the near-surface layers.

 \subsection{Torsional Oscillations}
 \label{sec:tor} 

One of the most striking results from helioseismology has been the
discovery that the so-called ``torsional oscillations'' -- which
modulate the observed pattern of surface differential rotation --
penetrate a substantial fraction of the convection zone. The surface
torsional oscillations were first observed by Howard and La Bonte
(1980). The observations showed bands of plasma at particular
latitudes rotating either slightly faster or slower (by a few per
cent) than the level expected from the smooth, underlying differential
rotation. What is more, the bands shifted position as the solar cycle
progressed, tracking toward the equator on a timescale that suggested
they carried the signature of the effects of the cycle.

%%%%%%%%%%%%%%%%%%%%%%%%%%%%%%%%%%%%%%%%%%%%%%%%%%%%%%%%%%%%%%%%%%%%%%%%%%%%%%%

\begin{figure*}
 \centerline
% {\includegraphics[width=0.9\textwidth,clip=]{fig1_48frames.ps}}
 {\includegraphics[width=0.9\textwidth,clip=]{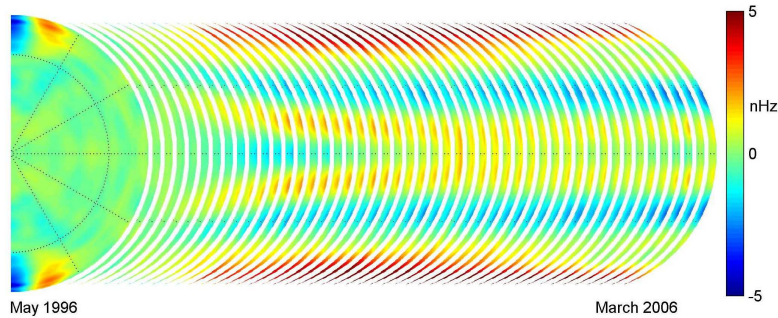}}

 \caption{Variations in the solar internal rotation, relative to the
 rotation at the epoch of solar minimum, as determined by analysis of
 MDI observations. (Courtesy of S.~Vorontsov.)}

 \label{fig:tors}
\end{figure*}

%%%%%%%%%%%%%%%%%%%%%%%%%%%%%%%%%%%%%%%%%%%%%%%%%%%%%%%%%%%%%%%%%%%%%%%%%%%%%%%

The modern, high-quality seismic observations (Figure~\ref{fig:tors})
reveal that these bands of flow are present within the convection zone
(Howe \etal, 2000a; Antia and Basu, 2000; 2001; Vorontsov \etal,
2002). Furthermore, an additional strong, poleward branch has been
revealed, which appears to penetrate the entire convection zone. The
amplitudes and phases of the signals show a systematic variation with
position in the convection zone (Howe \etal, 2005). The observed
behaviour of the flows is strongly suggestive of being a signature of
magnetic effects from the solar cycle.

An obvious candidate is the back-reaction of the magnetic field on the
solar plasma, via the Lorentz force. Lorentz force feedback was
proposed originally by Sch\"ussler (1981) and Yoshimura (1981) as a
means to explain the surface torsional oscillations. Later models
looked at the effect of the Lorentz force from small-scale magnetic
field on the turbulent Reynolds stresses (\textit{e.g.}, K\"uker,
R\"udiger, and Pipin, 1996). A thermal mechanism was also proposed by
Spruit (2003) to explain the low-latitude branch, having its origin in
small gradients of temperature caused by the magnetic
field. Incorporation of the Lorentz force feedback into dynamo models
(which are then termed ``dynamic'') can reproduce observed features of
the torsional oscillations (\textit{e.g.}, Covas, Moss, and Tavakol,
2005). Rempel (2007) has recently forced torsional oscillations in a
mean-field differential rotation model, which includes the effect of
the Lorentz force feedback in meridional planes. He found that while
the poleward propagating high-latitude branch could be explained by
Lorentz force feedback, or thermal driving, the low-latitude branch is
most likely not due to the Lorentz feedback, and probably has a
thermal origin.

Howe \etal\, (2006) conducted experiments using artificial data
containing migrating flows like those seen in the real
observations. Analysis of these artificial data suggests inferences
made on the depth of penetration, and the amplitude and phase, of the
solar torsional oscillations are likely to be real, and not artifacts
of the analysis. With the collection of more data, coupled to a better
understanding of the sub-surface torsional oscillations, it should be
possible to constrain the perturbations driving the flows
(\textit{e.g.}, Lanza, 2007). This might lead to the possibility of
obtaining indirect measures of the strength of the magnetic field with
depth in the convection zone (direct measurement of the field is not
yet possible).

 \subsection{The 1.3-yr Periodicities Near the Tachocline}
 \label{sec:13}

%%%%%%%%%%%%%%%%%%%%%%%%%%%%%%%%%%%%%%%%%%%%%%%%%%%%%%%%%%%%%%%%%%%%%%%%%%%%%%%

\begin{figure*}
 \centerline
% {\includegraphics[width=1.0\textwidth,clip=]{aurabase003.eps}}
 {\includegraphics[width=1.0\textwidth,clip=]{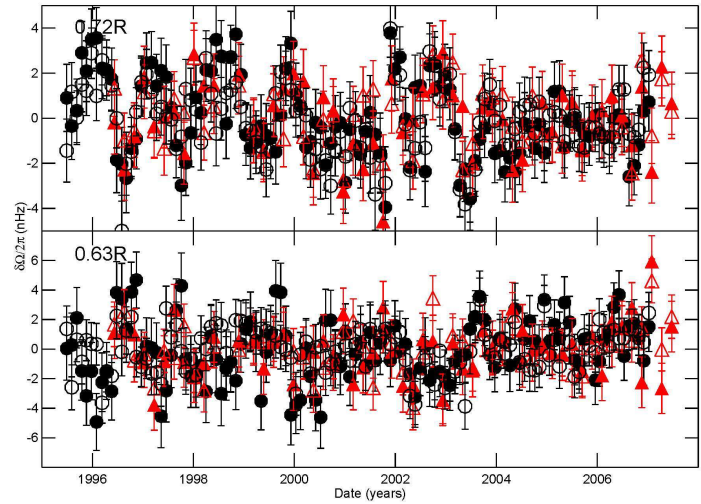}}

 \caption{The temporal variation of the rotation rate in the
  equatorial regions near the base of the convection zone at
  $0.72\,r/R_{\odot}$ (top) and at $0.63\,r/R_{\odot}$ (bottom) from
  GONG (circles) and MDI data (triangles). The average rotation rate
  has been subtracted. (Courtesy of R.~Howe.)}

 \label{fig:13year}
\end{figure*}

%%%%%%%%%%%%%%%%%%%%%%%%%%%%%%%%%%%%%%%%%%%%%%%%%%%%%%%%%%%%%%%%%%%%%%%%%%%%%%%

Claims that the rotation rate in the layers just above and below the
tachocline varies on a timescale of $\approx 1.3$\,years (Howe \etal,
2000b; see Figure~\ref{fig:13year}) remain controversial. When they
were first uncovered by the analysis, the changes appeared to be most
prominent in the low-latitude regions just above the base of the
convection zone. At the same time there were suggestions of variations
in anti-phase some $\approx$60\,000\,km deeper down, in the outer
parts of the radiative zone. The variations uncovered by the analysis
then all but disappeared when mid-latitude regions were tested, while
a periodic-looking signal of period closer to 1\,year was found when
attention was focused at latitude $60^{\rm o}$.

The result is controversial principally for two reasons. First,
independent analyses have failed to reveal the same quasi-periodic
variation of the rotation rate (Antia and Basu, 2001; 2004; Corbard
\etal, 2001). Second, the quasi-periodic signal appears to have all
but disappeared in more recent data, collected from 2004 onwards,
having also been absent over the period from $\approx 2000$ to
$\approx 2002$ (Howe, 2006). The intermittancy need not necessarily
imply the phenomenon is an artifact, and the claims continue to draw
considerable interest. The fact that the signals uncovered above and
below the tachocline are in anti-phase -- meaning as one region speeds
up the other slows down -- suggests that, if real, they may be
signatures of some form of angular momentum exchange between the
interior and envelope, mediated by the tachocline. Intriguingly, there
are also reports in the literature of quasi 1.3-year periodicities in
observations of sunspots and geomagnetic indices (see Howe, 2006, and
references therein).

  \section{The Observer's Holy Grail: G Modes and Very Low-Frequency $P$ Modes}
  \label{sec:gmodes}

As the title of this section suggests, the drive to detect the gravity
($g$) modes (and also the very low-frequency $p$ modes) has assumed an
added significance as time has passed. Detection of the $g$ modes
presents a major observational challenge, because the amplitudes of
the modes are predicted to be extremely weak at the photospheric
level. Early claims of detections of low-$\ell$ $g$ modes
(\textit{e.g.}, Delache and Scherrer, 1983; Fr\"ohlich and Delache,
1984) were to prove unfounded, and it has become increasingly apparent
that unambiguous detection of the modes will demand very long
datasets, excellent instrumental noise performance at low frequencies,
and ingenuity in both the observations (\textit{e.g.}, possibly from
new approaches to the observations) and the analysis.

Upper limits on the amplitudes of individual $g$ modes -- as set from
analysis of the long, high-quality modern datasets (\textit{e.g.},
Appourchaux \etal, 2000b; Gabriel \etal, 2002; Wachter \etal, 2003;
Turck-Chi\`eze \etal, 2004) -- are far superior (\textit{i.e.}, much
lower) than those set in the early analyses (where the datasets were
much shorter, and usually the quality of the data was inferior, to
contemporary levels). For some methods of analysis the limits are
approaching the level of $1\,\rm mm\,s^{-1}$ per mode. While
predictions of the $g$-mode amplitudes, based on the assumption of
stochastic excitation in the convection zone (\textit{e.g.}, Gough,
1985; Andersen, 1996; Kumar, Quataert, and Bahcall, 1996), may be
rather uncertain -- predictions for the same range in frequency can
differ by more than an order of magnitude -- it is worth noting that
some of the predictions are not too dissimilar from the current best
observational upper limits (\textit{e.g.}, see Elsworth \etal, 2006).

Early attempts to detect low-$\ell$ $g$ modes concentrated largely on
the very low-frequency asymptotic regime (\textit{e.g.}, see Provost
and Berthomieu, 1986), where the near constant spacing in period
offered potential advantages for detection algorithms. However,
amplitudes are expected to be appreciably larger in the
higher-frequency part of the $g$-mode spectrum. This is why the lack
of any convincing detections lower down shifted the focus of attempts
in the late 1990s toward the higher-frequency (non-asymptotic) range
(\textit{e.g.}, Appourchaux, 2003), where modes with mixed $g$ and $p$
characteristics are also expected. Searches were made, unsuccessfully,
for signatures of individual $g$ modes (although a few potential
candidates were claimed by Turck-Chi\`eze \etal, 2004).

Now, things have gone full circle. Searches in the very low-frequency
asymptotic regime are back in fashion. As noted above, analysis
methods can take advantage of the near-regular spacing in period, and
increase the effective signal-to-noise ratio by looking for the
cumulative effect of several $g$-mode overtones. Searches by Garc\'ia
\etal\, (2007) for the cumulative signature of $\ell=1$ $g$ modes in
almost 10\,years of GOLF data have yielded what may be regarded as the
first serious claim of a detection. The analysis technique has the
advantage that it is both elegant, and simple. The low-frequency part
of the frequency power spectrum is first pre-whitened. The
pre-whitened part is then presented in the form power spectral density
against period, and its periodogram computed. The cumulative signature
of the $\ell=1$ overtones is manifested as a peak in the periodogram,
at a period corresponding to the period spacing between the overtones.

Garc\'ia \etal\, claim a statistically significant peak in the GOLF
periodogram, which lies at roughly the period predicted by the
standard solar models. Is this really the signature of $\ell=1$ $g$
modes?  Experiments performed by the authors with artificial data
suggest the individual mode peaks must have widths in the frequency
power spectrum that are commensurate with damping times of several
months. Excitation of gravity waves at the base of the convection zone
might lead to the comparatively heavy damping implied by these values
(Dintrans \etal, 2005), although the theoretical work is not yet
sufficiently developed to enable accurate damping-rate predictions to
be made for the global low-$\ell$ $g$ modes.

The validity of conclusions drawn on all searches for low-frequency
modes rest on a robust and proper use of statistics. Some methods
assess the likelihood that prominent features are part of a broad-band
noise source, the so-called H0 hypothesis; while others test against
the likelihood that signal (\textit{e.g.}, a sine wave or a damped
wave) is buried in broad-band noise, the so-called H1 hypothesis (see
Appourchaux, 2003, and Chaplin \etal, 2004, for brief discussions on
low-frequency detection methods).  Some of the tests bring more prior
information to bear than others: for example, tests are often
predicated on the assumption that the $g$ modes are very lightly
damped, like the low-frequency $p$ modes, meaning individual
components will appear as spikes in the frequency power spectrum. It
is also important to recognise that quoted likelihoods depend on the
question being asked of the data. For example, does one flag a
prominent spike as a candidate mode if it has less than a 1\%
likelihood of appearing by chance in a range of $10\,\rm \mu Hz$ of
the spectrum? Or does one perhaps demand that it has less than a 1\%
chance of appearing in a range of $100\,\rm \mu Hz$ of the spectrum?
Quoting a 1\% likelihood in these two cases means two different things
(the second criterion being a more demanding limit).  And should one
fold in the fact that the number of bins in a fixed range of frequency
increases as more data are collected on the Sun?

The use of prior information, and the need to fix \textit{a priori}
choices for hypothesis testing (which has an element of subjectivity),
suggests that a Bayesian (and not a frequentist) approach is the best
route to assessing the likelihoods associated with searches for the
$g$ modes.  This is the approach currently advocated by the Phoebus
collaboration, which is leading the way on development of analysis
techniques in this area (Appourchaux, 2008).

The application of low-frequency detection algorithms is now yielding
multiple detections on $p$ modes at frequencies below $1000\,\rm \mu
Hz$, though at $\ell \ge 4$ (\textit{e.g.}, see Salabert, Leibacher,
and Appourchaux, 2008). The lowest frequency detection at $\ell \le 3$
-- which has received independent confirmation by different analyses
of more than one dataset -- is the $\ell=0$, $n=6$ overtone at
$\approx 973\,\rm \mu Hz$ (\textit{e.g.}, see Garc\'ia \etal, 2001;
Broomhall \etal, 2007).

 \section{Driving and Damping the Modes}
 \label{sec:excite}

The first calculations of the excitation rates of the $p$ modes
suggested the modes were unstable (\textit{e.g.}, see the discussion
in Christensen-Dalsgaard, 2004). We now know they are in fact stable,
being stochastically excited and intrinsically damped by the
convection (see Houdek, 2006, for a recent review). Theoretical
modelling of global excitation and damping based on an analytical (or
semi-analytical) approach (\textit{e.g.}, Houdek \etal, 1999; Dupret
\etal, 2004; Samadi \etal, 2005) can give predictions of two
independent quantities: the damping rates ($\eta$), and acoustic
powers ($P$) (the latter corresponding to the rates at which energy is
pumped into, and then dissipated by, the modes). It is therefore
incumbent on the observers to provide as accurate and precise measures
of these parameters as possible.

The parameters that are usually extracted directly by the observers
are the widths ($\Delta$) and heights (maximum power spectral
densities) ($H$) of the peaks in the frequency power spectrum. The
linear damping constants ($\eta$) are related to the peak widths via:
 \begin{equation}
 \Delta = \eta / \pi.
 \label{eq:damp}
 \end{equation}
If observations are of sufficient length to resolve mode peaks in the
frequency power spectrum, the observed heights of the mode peaks are
given by:
 \begin{equation}
 H = \frac{2V^2}{\pi \Delta} = \frac{P}{\eta^2 I},
 \label{eq:height}
 \end{equation}
where the $V$ are mode amplitudes (written here for Doppler velocity)
and the $I$ are the mode inertias. There has recently been a shift
toward making comparisons of observational and theoretical mode
amplitudes using the $H$ (\textit{e.g.}, Chaplin \etal, 2005; Belkacem
\etal, 2006a), as opposed to the $V$ (or $V^2$) as had previously been
the practice. The $H$ is after all what one ``sees'' for the vast
majority of modes in the frequency power spectrum (\textit{i.e.},
provided they are well resolved).

Measurement of the acoustic powers ($P$) from the peak-bagging
estimates of $H$ and $\Delta$ is fraught with potential
pit-falls. Equations~(\ref{eq:damp}) and~(\ref{eq:height}) imply that:
 \begin{equation}
 P = \pi^2 I H \Delta^2.
 \label{eq:pobs}
 \end{equation}
There is a strong anti-correlation of the fitted $H$ and $\Delta$ in
fits made to peaks in the frequency power spectrum. The effect cancels
when estimates of $V^2$ are sought, since $V^2 \propto H \Delta$; but
estimates of $P$ contain another factor of $\Delta$. Some other means
of estimating the damping, that is much less strongly correlated with
$H$, would offer a way around the problem.

The appearance of the mode inertia ($I$) in Equation~(\ref{eq:pobs})
gives rise to further complications. Because different instruments
show different Doppler velocity responses with height in the
photosphere, the $I$ are instrument (\textit{i.e.}, observation)
dependent.  Baudin \etal, (2005) have demonstrated the importance of
attempting to correct for this effect.  Without proper normalization
between results of different instruments, differences in estimates of
$P$ arise.

The frequency dependence of the acoustic powers, $P$, is a
particularly important diagnostic. The most recent comparisons of
theory and observation have shown good agreement over the main part of
the low-$\ell$ mode spectrum (\textit{e.g.}, Chaplin \etal, 2005;
Belkacem \etal, 2006a, b). Samadi \etal, (2007) have also found that
absolute predictions of $P$ by three-dimensional numerical simulations
tend to be lower than the $P$ given by the semi-analytical
models. With regard to the damping, the comparison, by Chaplin \etal\,
(2005) of the observed and theoretical $H$ of the radial modes has
demonstrated much more clearly than before the shortcomings in the
theoretical computations of the low-frequency damping rates.

An interesting question for the analytical models concerns the
description of the temporal behaviour of the dynamics of the
small-scale turbulence. A Gaussian or Lorentzian function is usually
adopted. Chaplin \etal, (2005) found that when they adopted a
Lorentzian description the predicted $H$ of the low-frequency modes
severely overestimated observed values. They concluded that a Gaussian
description gave a much better match to the observations. The results
of Samadi \etal\, (2003, 2007) in contrast tend to favour the
Lorentzian description. Results from numerical simulations indicate
that neither description is strictly correct (Georgobiani, Stein, and
Nordlund, 2006): variation in the behaviour of the small-scale
turbulence is observed with both temporal frequency and depth in the
convection zone.

The two sources of excitation are the fluctuating turbulent pressure
(Reynolds stresses) and gas pressure. In the numerical simulations
(\textit{e.g.}, Stein \etal, 2004) there is some cancellation between
the sources. This cancellation is not shown by the analytical models,
which may explain why the analytical models overestimate the $p$-mode
amplitudes of stars hotter than the Sun (see discussion in Houdek,
2006).

  \section{The Road Ahead}
  \label{sec:ahead}

Global helioseismic studies continue to make great progress, but many
outstanding questions and challenges remain. If there is one thing
that 30\,years of global helioseismology has taught us, it is that
there are two highly desirable requirements on the observational data:
\textit{i}) that they should provide long-term, high-duty-cycle
monitoring of modes from low to high $\ell$; and \textit{ii}) that
they should offer observational redundancy, and complementary data.

Requirement \textit{i} enables us to use the global $p$ modes to
``sound'' the solar activity cycle, \textit{i.e.}, to monitor in
detail the temporal behaviour of the seismic Sun on timescales
commensurate with the cycle period, and preferably longer to
facilitate comparisons of one cycle with another. It also enables us
to obtain extremely precise and accurate estimates of fundamental mode
parameters -- long datasets are vital for detection of the very
low-frequency modes -- and to measure other parameters that would not
otherwise be determined robustly (\textit{e.g.}, multiplet frequency
asymmetries in low-$\ell$ modes). We are therefore in a position to be
able to make precise and accurate inference on the internal structure
and dynamics (both the time-averaged properties, and the properties as
a function of time).

Requirement \textit{ii} enables us to confirm the solar origins of
subtle, but potentially important, phenomena in the data. For example,
detection of weak very low-frequency modes in two or more
contemporaneous datasets significantly lowers the probability of a
false detection having been made.  Complementary Doppler-velocity and
intensity observations, and observations by Doppler velocity
instruments in different atmospheric absorption lines, create
opportunities for studies of the physics of the photosphere, studies
which can in turn be used to obtain more accurate estimates of mode
frequencies from better understanding and modelling of the peak
asymmetry.

To fully exploit the potential science benefits that global
helioseismology has to offer, we need continuation of operations of
the two main ground-based networks, GONG and BiSON. As new science
results drive the need for different data products, the ground-based
networks are in a position to implement ``responsive'' changes to
their instrumentation (we come back to the issue of new observational
requirements later). In the post-SOHO era, BiSON will continue to
provide, and will then be the only bespoke source of, high-quality
low-$\ell$ data from its Sun-as-a-star observations. Continuation of
GONG, in its current multi-site configuration, would provide
high-quality, high-duty-cycle resolved-Sun products, particularly on
higher-$\ell$ modes, to go alongside the HMI resolved-Sun data (due
for launch on SDO in early 2009).

It is important to remember that to make optimal use of the low-$\ell$
modes for probing the solar core we need contemporaneous medium- and
high-$\ell$ data It is worth stressing the important r\^ole the
high-$\ell$ modes can play in this regard, in that they can be used to
constrain the hard-to-model near-surface layers, thereby cleaning
things up for more accurate inference on the structure deeper
down. However, reliable measurement of the high-$\ell$ frequencies
presents something of a challenge, because of the sensitivity of the
frequencies to instrumental effects (\textit{e.g.}, see Korzennik,
Rabello-Soares, and Schou, 2004; Rabello-Soares, Korzennik, and Schou,
2007).

In order to further improve the accuracy of the inversions we must
continue studies into optimizing combinations of frequencies from
different instruments (\textit{e.g.}, the low-$\ell$ Sun-as-a-star
BiSON and GOLF data with the resolved-Sun MDI and GONG, and in the
future the HMI, data). As datasets get longer, and quality improves,
so new subtle effects come to light that must be properly allowed for
when the datasets are analyzed. In the last few years we have
developed a much better understanding of the underlying frequency bias
between resolved-Sun and Sun-as-a-star frequencies. But more work is
needed. New instrument combinations inevitably present their own
unique problems.

Bias comes not only from instrumental effects, but also from the
analysis pipelines (\textit{e.g.}, see Schou \etal, 2002; Basu \etal,
2003). Hare-and-hounds exercises on realistic artificial data are a
valuable tool for uncovering, and understanding, such effects. The
solarFLAG group is currently concluding a second round of
hare-and-hounds exercises testing peak-bagging on low-$\ell$ modes in
Sun-as-a-star data (see Chaplin \etal, 2006 for results on the first
round of exercises). Significant improvements to the peak-bagging at
medium and high $\ell$ are being made by Jefferies and Vorontsov
(2004) and Korzennik (2005). The approach of Jefferies and Vorontsov
-- parametric modelling of the spectrum using a small number of free
parameters -- is novel. The approach has the potential (and indeed the
ultimate aim) to ``remove the intermediary'' -- by which here we mean
estimation and subsequent use of individual mode parameters, like the
frequencies -- to instead give the desired constraints on models of
the internal structure by maximizing the likelihood of the solar model
parameters directly on the frequency spectra. The importance of
detailed work on the peak-bagging codes and philosophies should never
be overlooked.

Before we finish, let us go back briefly to the observations. New
observations on the modes in intensity (from low to medium $\ell$)
will be provided by the SODISM and PREMOS instruments on PICARD (due
for launch in early 2009). While a prototype next-generation GOLF
instrument (GOLF-NG) is about to begin ground-based trials in
Tenerife. The SODISM and GOLF-NG instruments are testing new
techniques in the observations, which will hopefully increase the
likelihood of detecting the low-frequency $g$ modes. SODISM will look
to the solar limb to maximize the signal-to-background ratio in the
$g$ modes. GOLF-NG will make simultaneous observations at different
heights in the solar atmosphere. The aim will be to take advantage of
changes in the coherence of the granulation signal with height to beat
down the solar noise background. Extension of this capability to
resolved-Sun observations is clearly a desirable goal (Hill, 2008).

So, looking to the future, we must advocate strongly for the
continuation of unbroken, high-duty-cycle ``seismic monitoring'' of
the Sun, at low, medium and high $\ell$. There are exciting challenges
for the observations and analysis: for example, to detect and identify
individual low-$\ell$ $g$ modes, and to measure their properties, in
particular the frequencies and frequency splittings; to use long-term
monitoring of the global $p$ modes to detect evidence of long-term
secular change in the Sun's seismic properties; to use the long-term
monitoring to enable comparisons of different 11-year activity and
22-year magnetic cycles, using low, medium and high-$\ell$ modes; and
to be able to fully isolate, and then subtract from the mode
frequencies, the influence of the near-surface layers, and to thereby
reveal a ``cleaner'' picture of the structure of the deep interior.

Some key questions for global helioseismology to address include: what
is the solar composition as a function of radius in the interior, and
is the solar abundance problem a problem in the spectroscopic
abundance determinations, or a problem in the standard solar models
(\textit{e.g.}, is there something missing from the models)?  What is
the strength of the magnetic field at the tachocline, and in the
convection zone, and what are the implications for solar dynamo
models?  What is the rotation profile as a function of radius in the
solar core, and what are the implications for models of the dynamic
evolution of Sun-like stars?

\acknowledgements

The authors are very grateful to R.~Howe and S.~Vorontsov for
providing figures and to G.~Houdek for useful discussions.

\end{article}
\end{document}